\documentclass[twocolumn, superscriptaddress, prl, floatfix]{revtex4-1}
\usepackage{amssymb, amsmath, graphicx, natbib}
\usepackage{xcolor}  

\begin{document}

\title{Lipid Bilayer Hydrodynamic Drag}

\author{Philip E. Jahl}
\affiliation{Department of Physics and Materials Science Institute, The University of Oregon, Eugene, OR, 97401}

\author{Raghuveer Parthasarathy}
\affiliation{Department of Physics and Materials Science Institute, The University of Oregon, Eugene, OR, 97401}

\date{\today}

\begin{abstract}
The hydrodynamic drag at a lipid bilayer surface determines in part the flow properties of suspensions of cells and liposomes. Given the fluidity of lipid bilayers, it is not obvious {\em a priori} whether solid-like no-slip,  liquid-like no-stress, or intermediate boundary conditions apply at the water-bilayer interface. Though no-slip conditions have been widely assumed for many decades, this fundamental aspect of membrane rheology has, to our knowledge, never been directly measured for free bilayers. We applied light sheet fluorescence microscopy to image freely diffusing phospholipid vesicles and determined the hydrodynamic drag coefficient $C \pi \eta R$, where $\eta$ is the external fluid viscosity, $R$ is the vesicle radius, and the dimensionless $C$ characterizes the flow boundary condition. We find that  $C = 5.92 \pm 0.13$ (stat.) $\pm$ 0.16 (syst.), matching the theoretical value of $C=6$ for a no-slip boundary and far from the $C=4$ value for a zero shear stress boundary.
\end{abstract}

\maketitle

Interfaces between lipid bilayers and aqueous solutions are present in countless environments both natural, such as at cell and organelle membranes, and artificial, such as in suspensions of liposome-encapsulated drugs. The hydrodynamics of bilayers and bilayer-bound objects are therefore of considerable interest \cite{Stachowiak2006, Brown2011, Tabaei2016, Vogele2018, Cicuta2007, Hormel2014, Hormel2015, Thoms2017}. In particular, the rheology of red blood cells {\em in vivo} \cite{Mendez2018} and suspensions of cells and liposomes {\em in vitro} \cite{Lee1993, He2010} depends directly on the nature of the flow boundary condition of the bilayer / water interface. It has been widely assumed throughout work spanning many decades that this interface is described by the no-slip condition characteristic of solid-liquid interfaces \cite{Tabaei2016, Mason1978, Foo2004, Johnson1971, Kraus1996, Waugh1982}. However, lipid bilayers are Newtonian fluids \cite{Hormel2015, Cicuta2007, Petrov2012}, and it has been speculated that their aqueous interfaces may therefore behave more like no-shear-stress boundaries, or intermediate between solid-like and liquid-like extremes \cite{Stachowiak2009}. Remarkably, despite its fundamental importance and widespread applicability, we are aware of almost no measurements of the flow boundary conditions at lipid bilayer surfaces. One study was able to use a dynamic surface force apparatus to probe lipid bilayers supported on solid surfaces, reporting no-slip flow conditions \cite{Cross2006}. However, the presence of a solid support is well known to alter membrane hydrodynamics, reducing lipid diffusion by roughly an order of magnitude compared to free bilayers and inhibiting large-scale spatial organization \cite{Kaizuka2004, Stottrup2005}. To the best of our knowledge, determining the flow boundary conditions at free lipid bilayers remains, surprisingly, an open problem.

In principle, the flow boundary condition of an object can easily be determined from measurements of its Brownian Motion. For a sphere of radius $R$, the diffusion coefficient $D$ is given by the the Stokes-Einstein relation
\begin{equation}
D = \frac{k_B T}{C \pi \eta R},
\label{EqStokesEinstein}
\end{equation}
where $k_B$ is Boltzmann's constant, $T$ is the temperature, $\eta$ is the external fluid viscosity, and $C$ is a dimensionless constant that characterizes the boundary condition. It is well known that $C=6$ corresponds to a no-slip condition, as is the case for ideal solids, and $C=4$ corresponds to an interface with no shear stress, as is the case for a liquid sphere of zero viscosity. More generally, for a sphere of viscosity $\eta_{internal}$ in an external liquid of viscosity $\eta_{external}$, with 
\begin{equation}
\lambda = \frac{\eta_{internal}}{\eta_{external}},
\end{equation}
the boundary constant is \cite{Deen1998, Hadamard1911, Rybczynski1911}:
\begin{equation}
C = 4\frac{3\lambda+2}{2\lambda+2}
\end{equation}

In practice, determining $C$ by measuring $D$ is non-trivial due to potential hydrodynamic influence from nearby surfaces such as container walls, and the requirement of high-precision determination of the object's positions and radius.

We surmount these challenges by applying light sheet fluorescence microscopy \cite{Ntziachristos2010, Loftus2013, Taormina2012, Keller2008} together with fast, accurate tracking techniques \cite{Parthasarathy2012} to characterize the diffusive motion of spherical phospholipid vesicles. Light sheet fluorescence microscopy provides optical sectioning of three-dimensional samples, enabling the imaging of vesicles hundreds of microns (tens of vesicle diameters) away from the walls of the imaging chamber \cite{Loftus2013}. We verified our methodology by characterizing the diffusive motion of solid microspheres in an aqueous medium and water droplets in benzyl alcohol, as detailed below, which gives $ C = 6.28 \pm 0.15$ and $C = 4.36 \pm 0.28$, respectively, consistent with theoretical expectations. We then characterized lipid vesicles composed primarily of the common phospholipid DOPC (1,2-dioleoyl-sn-glycero-3-phosphocholine), determining that $C = 5.92 \pm 0.13$ (stat.) $\pm$ $0.16$ (syst.). This establishes that the bilayer/water interface is well described by a no-slip boundary condition.

\section{Methods}

\begin{figure}[h]
\includegraphics[scale=.4]{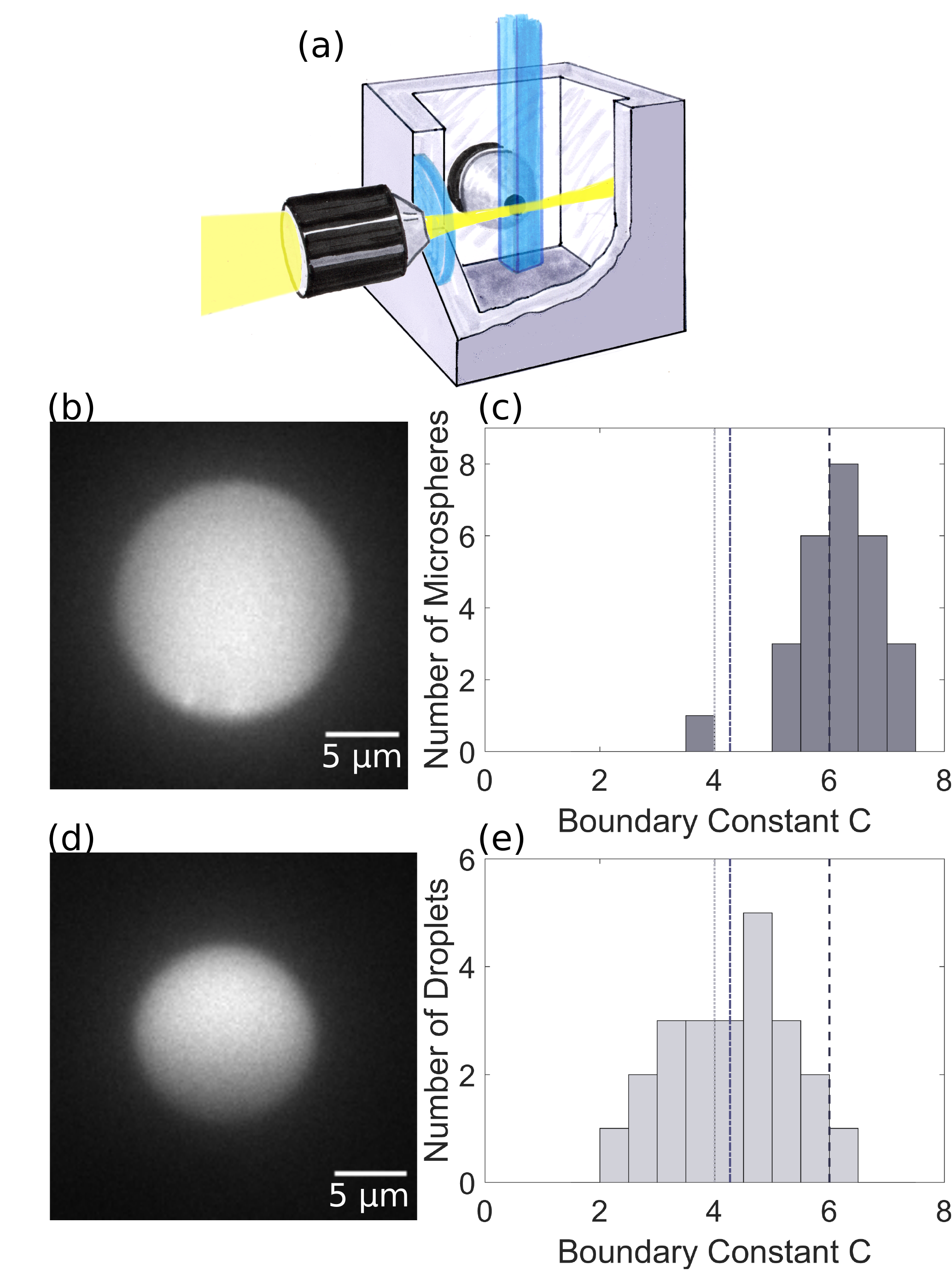}
\caption{(a) A schematic of the setup for light sheet fluorescence microscopy of vesicle diffusion. The excitation laser is shown entering the sample chamber from the left as it is focused into a thin sheet. Light emitted by the sample is collected by the objective lens shown behind the cuvette. (b) A typical light sheet fluorescence image of the central plane of a 15 $\mu$m diameter polystyrene microsphere. (c) Histogram of $C$ for 29 particles, giving a mean $\pm$ standard error of $C = 6.28 \pm 0.15$.   (d) A typical light sheet fluorescence image of the central plane of a fluorescein-dyed water droplet in benzyl alcohol. (e) Histogram of $C$ values for 25 droplets, giving a mean $\pm$ standard error of $C = 4.36 \pm 0.29$. In (b) and (d), the dashed-dotted line indicates the theoretical value of $C=4.27$ for water in benzyl alcohol and the dashed and dotted lines respectively indicate the theoretical values of $C = 6$ and $C=4$ for ideal no-slip and slip boundary conditions.}
\label{fig: Bead and Oil Results}
\end{figure}

We performed light sheet fluorescence microscopy using a home-built instrument that closely follows the design of Keller et al. \cite{Keller2008} and is described in detail in \cite{Taormina2012}. In brief, excitation light was provided by a 488 nm laser with an output power of 50 mW, which was scanned by a galvanometer mirror and focused by an objective lens to form a sheet in the sample chamber. The minimum thickness of the sheet was $\approx 3$ $\mu$m, extending over a lateral extent (Rayleigh length) of $\approx 100$ $\mu$m. Images were captured through a $40\times$ 1.0 NA Plan-apo objective lens (Zeiss) perpendicular to the excitation plane and recorded with a 5.5 Mpixel sCMOS camera (pco.edge, Cooke Corp.). A schematic of the setup is shown in Figure \ref{fig: Bead and Oil Results}(a).

Suspensions of either lipid vesicles or polystyrene microspheres in 0.1 Molar sucrose, or deionized water droplets in benzyl alcohol, were placed in a square cross-section glass cuvette (Starna Cells, part number 3-3.45-SOG-3), which was mounted to a movable translation stage and inserted in the light sheet microscopy chamber. The distance between imaged objects and the cuvette walls was several hundred micrometers.

For beads, droplets, and vesicles the optical plane intersecting the sphere center (the ``equatorial'' plane) was readily evident due to a lack of out-of-focus light outside the bright ring or disk, due to the few micron sheet thickness. This is illustrated in Supplemental Movie 1, which shows a three-dimensional scan through a lipid vesicle.

To assess the accuracy of our methods, we examined diffusion of objects with well known flow boundary conditions: solid polystyrene microspheres and deionized water droplets suspended in benzyl alcohol.

The microspheres were FITC-labeled polystyrene beads of nominal  diameter 15.45 $\pm$ 0.70 $\mu$m (mean $\pm$ standard deviation; Bangs Laboratory, part number FSDG009). Light sheet fluorescence images were captured for durations of 15 seconds at 33.33 frames per second. A typical image is shown in Figure \ref{fig: Bead and Oil Results}(b). For the first image, we determine the particle center using the radial symmetry-based algorithm described in \cite{Parthasarathy2012}, which provides rapid localization with accuracy close to theoretical limits. In brief, the center is calculated as the point that minimizes total distance to lines derived from intensity gradients throughout the image. The original algorithm was modified to only weight intensity gradients from the vicinity of the bead, limiting the effects of noise outside the particle, such as fluorescent debris or light from other beads. Each remaining image was cross correlated with the previous image and the original radial symmetry-based algorithm was applied to the cross-correlation to determine the shift between each frame. From the particle positions, we determine the diffusion coefficient $D$ using the covariance based estimator described by Vestergaard et. al. \cite{Vestergaard2014}. Not only does this provide greater accuracy than, for example, linear fits of mean-squared-displacements to time intervals, but it also provides robust estimates of localization accuracy and goodness-of-fit to a random walk model \cite{Vestergaard2014} that we make use of in assessing vesicle data below. We experimentally determined the radius of each bead by Hough transformation following edge detection \cite{Illingworth1987}, which produces from each input image a series of output images corresponding to each possible radius candidate, with the true object radius giving a bright, compact transform image. Across all beads, this gives an average diameter of 15.06 $\pm$ 0.41 $\mu$m (mean $\pm$ standard deviation), consistent with the nominal value of 15.45 $\pm$ 0.70 $\mu$m. Using the measured radius, the literature value for the viscosity of deionized water, and the ambient temperature $T = 293$ K, we use Eq. \ref{EqStokesEinstein} to determine $C$ for each microsphere. We show the histogram of $C$ values for $N=29$ microspheres in Figure \ref{fig: Bead and Oil Results}(b). The mean $\pm$ standard error is  $C = 6.28 \pm 0.15$, slightly higher than but consistent with the expected $C=6$ for solid particles. (Using the nominal rather than the measured microsphere radius gives $C = 6.12$, matching the theoretical value within uncertainties. The microsphere edges are less well-defined than those of the vesicles; in the latter case, we note explicitly the uncertainty in $R$ below.)

The assessment of liquid droplets is similar. The droplets were deionized water dyed with 75 mg/ml fluorescein, suspended in benzyl alcohol, formed into an emulsion by vigorous shaking. Fluorescein has very low solubility in water, but nonetheless preferentially labels the aqueous phase. Benzyl alcohol was chosen because its density, 1.045 g/ml, is  similar to that of water, limiting gravity-induced drift. The viscosity of benzyl alcohol is $\eta = 6.29\times 10^{-3}$ Pa$\cdot$s \cite{Chen2012}, and so the expected $C = 4.27$. The droplets were imaged for 60 seconds at 8.33 frames per second. A typical image is shown in Figure \ref{fig: Bead and Oil Results}(c). Using the same procedures described above for center positions and radii, we determined $C$ for each water droplet. Figure \ref{fig: Bead and Oil Results}(d) shows the histogram of $C$ values for $N=25$ droplets. The mean $\pm$ standard error is $C = 4.36 \pm 0.29$, consistent with the expected value. Notably, $C$ increased with time for these droplets, likely due to adsorption of fluorescein to the boundary. The $C$ value stated was determined from data within 20 minutes of emulsion preparation.

The lipid vesicles we examined were composed of 94\% DOPC (1,2-dioleoyl-sn-glycero-3-phosphocholine), and 6\% NBD-PE (1,2-dipalmitoyl-sn-glyercero-3-phophoethanolamine-N-(7-nitro-2-1,3-benzoxadiazol-4-yl) (Avanti Polar Lipids). Phosphatidylcholine lipids are a major constituent of cellular membranes. The headgroup-conjugated NBD probe has an isotropic orientation relative to the lipid bilayer plane (unlike, for example, probes such as Texas Red \cite{Groves2008}), and so provides vesicle images of symmetric fluorescence under polarized laser excitation. The vesicles had radii between 3.1 and 25.7 $\mu$m with a mean of 12.2 $\mu$m and a standard deviation of 4.0 $\mu$m. Vesicles were imaged for 15 seconds at 33.33 frames per second. The vesicles were created by electroformation, as in \cite{Veatch2007}. In brief, the desired lipids were dried on glass slides with an indium tin oxide coating, hydrated with a 0.1 M sucrose solution, and subjected to an oscillating electric field to stimulate vesicle formation. Vesicles were added to a sample cuvette containing 0.1 M sucrose so that the interior and exterior of the vesicles would be matched in density and osmolarity. We use the literature value for the viscosity of 0.1 M sucrose, $\eta = 1.095\times 10^{-3}$ Pa$\cdot$s \cite{NIST1958}, in our analyses. Low levels of drift were present in the experiments, possibly due to convection and imperfect density matching. We therefore subtracted the best-fit linear trajectory (i.e. constant velocity) from each vesicle trajectory, and used only the horizontal component of trajectories. Rare frame-to-frame displacements more than three standard deviations from the mean ($< 0.5\%$ of frames) appeared to indicate large-scale instrument vibrations, and trajectories were analyzed piecewise around such points.

We assessed the accuracy of center- and radius-finding algorithms for vesicle images by applying them to simulated images of bright rings with a range of radii and signal-to-noise ratios (SNRs), mimicking the form of the vesicle images, as in Ref. \cite{Loftus2013} and similar to the ring images in Ref. \cite{Parthasarathy2012}. In brief, a high-resolution image of a thin annulus was convolved with the detection point-spread function (PSF), pixelated, and subjected to Poisson-distributed noise. We used the theoretical PSF of the emission wavelength and numerical aperture, which has a full-width at half-maximum of 0.19 $\mu$ m. Because of the occasional presence of lipid matter in and around the vesicle edge, the radial center-finding algorithm is weighted with a hyperbolic tangent function centered around the ring of the vesicle so as to only use the intensity gradient from the edge of the vesicle. From analysis of simulated images, the radial-symmetry-based localization gives an estimated localization error of $\approx 3$ nm. Independently, the localization error estimated from vesicle trajectories by the method of Vestergaard et al.\cite{Vestergaard2014} is $\approx 10$ nm. The vesicle radius  is determined by Hough transformation, as in the bead and droplet image analysis. The estimated uncertainty in $R$ from the standard deviation of simulated images at the appropriate SNR is $\approx 0.005$ $\mu$m. More significant, however, is the standard deviation in $R$ over the course of an image series, due for example to changes in position relative to the sheet plane. This is $\approx 0.03$ $\mu$m, which is small compared to the typical 10 $\mu$m vesicle radii and which contributes negligibly to the overall uncertainty in $C$. From vesicle position data, we calculated $D$ and $C$ as described above.

\section{Results}

\begin{figure}[bht]
	\includegraphics[scale=.4, angle=270]{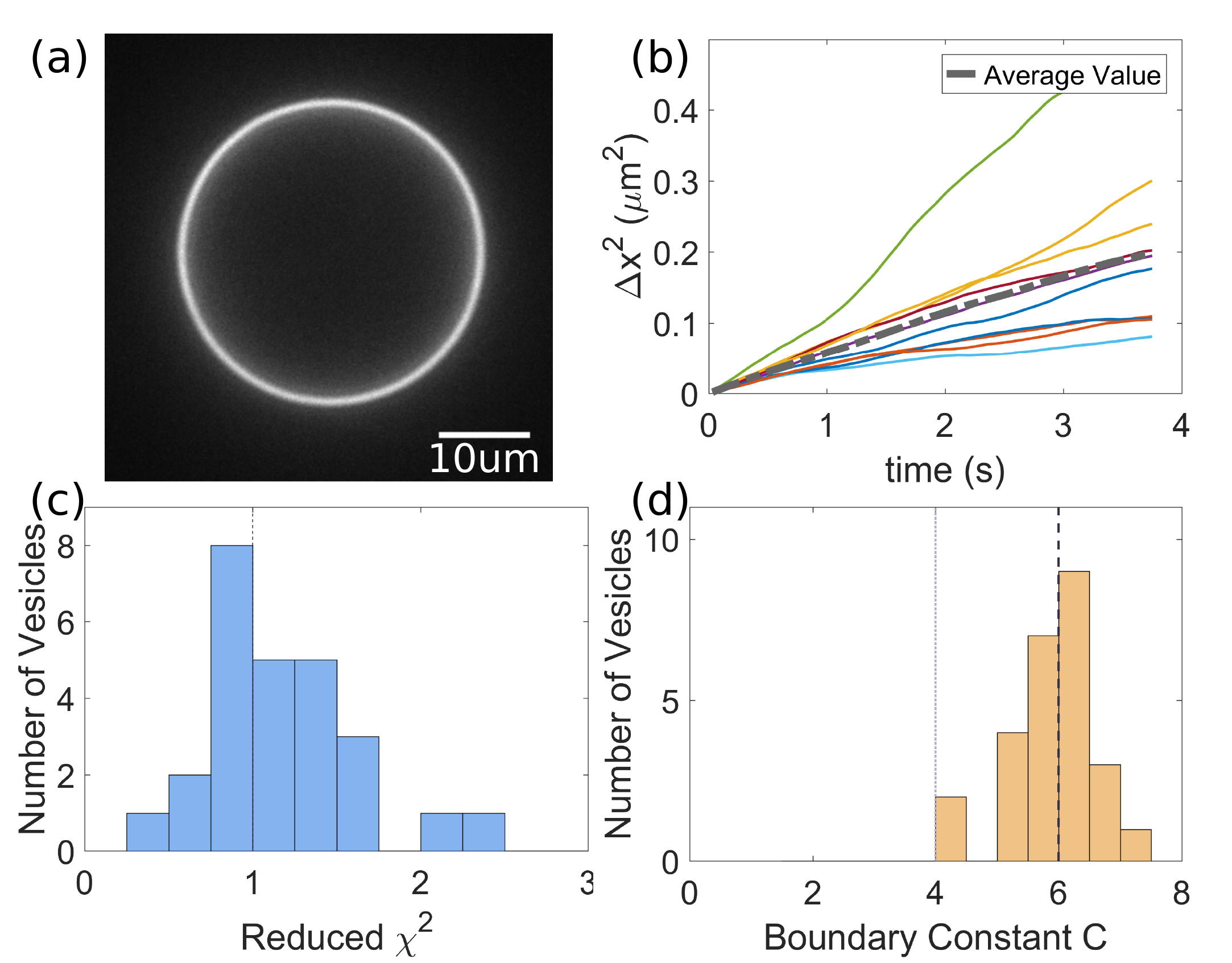}
	\caption{(a) A typical light sheet fluorescence image of the central plane of a DOPC vesicle. (b) Mean-square-displacements from 10 randomly chosen vesicle trajectories (colored lines), along with their average (dashed gray line). (c) Histogram of the reduced $\chi^2$ values for each vesicle; $\chi^2=1$ indicates purely diffusive motion. (d) Histogram of $C$ for 26 vesicles giving a mean $\pm$ standard error of $C = 5.92$ $\pm$ $0.13$. The dashed and dotted lines indicate the theoretical values of $C = 6$ and $C=4$ for ideal no-slip and slip boundary conditions, respectively.}
\label{fig: Vesicle Results}
\end{figure}

Light sheet fluorescence microscopy provides clear images of lipid vesicles. A typical example is shown in Figure \ref{fig: Vesicle Results}(a). Assessment of images as described above provides vesicle positions, radii, diffusion coefficients ($D$), and the flow boundary constant ($C$), along with their uncertainties. We provide the complete set of positions, radii, diffusion coefficients, and boundary coefficient values for every microsphere, water droplet, and lipid vesicle examined as Supplemental File.

In Figure \ref{fig: Vesicle Results}(b) we show ten randomly chosen examples of vesicle mean-squared-displacement(MSD) (colored lines) as a function of lag time together with their average (dashed gray line). The linearity of the MSD curve is indicative of free diffusion. A more robust assessment of the Brownian character of vesicle motion comes from the goodness of fit calculation provided by the covariance based estimator of $D$, which gives a reduced $\chi^2$ value whose value should be $\approx 1$ for a model of pure diffusion. A histogram of the measured $\chi^2$ values is shown in Figure \ref{fig: Vesicle Results}(c), and is consistent with simple Brownian diffusion. 

As described in Methods, we use measurements of vesicle radii and diffusion coefficients to determine the flow boundary condition constant $C$. The histogram of $C$ for $N=26$ vesicles is shown in Figure \ref{fig: Vesicle Results}(d).  The mean $\pm$ standard error is $C = 5.92 \pm 0.13$. To estimate possible systematic error, we assume that the polystyrene microspheres for which we calculated $C$ (see Methods) are ideal hard spheres. The standard deviation of the beads' $C$ value would therefore be the spread inherent in our methodology. In order to account for this uncertainty we can compute a systematic standard error by dividing the standard deviation of the beads' $C$ by the square root of the number of vesicles. This gives us a constant for vesicles of $C = 5.92$ $\pm$ 0.13 (stat.) $\pm$ 0.16 (syst.).

Our measurements show that at least over micrometer length scales and millisecond-to-second timescales, lipid bilayers are very well described by the no-slip flow boundary conditions that characterize solid surfaces. Perhaps reassuringly, the standard assumption that is ubiquitous in treatments of liposome hydrodynamics is well supported. It is perhaps surprising that fluid lipid membranes behave as solid surfaces with respect to flow through an external liquid, but it is consistent with the membranes having a large viscosity. Hydrodynamics is qualitatively different in two and three dimensions, and two-dimensional viscosity is dimensionally distinct from its three-dimensional counterpart. Nonetheless, one can divide the viscosity of a thin two dimensional fluid by its thickness to give a value that allows rough comparison to the viscosity of bulk three-dimensional liquids. For a lipid bilayer, the viscosity is on the order of $10^{-9}$ Pa$\cdot$s$\cdot$m \cite{Hormel2015, Stanich2013, HonerkampSmith2013}, and the thickness on the order of $10^{-9}$ m, giving a rough effective viscosity $\sim 1$ Pa$\cdot$s, corresponding to $\lambda \sim 1000$ and $C \approx 6$. We caution, however, that a lipid vesicle does not simply behave as a bulk $\sim 1$ Pa$\cdot$s droplet; external flow, for example, easily induces stresses in the thin membrane that generate interior flows \cite{HonerkampSmith2013}.

Our method shows that a conceptually simple imaging-based approach can provide precision measurements of microscale fluid properties. We expect that this can be extended to, for example, fluctuating membranes  driven by either temperature \cite{Loftus2013, Henriksen2004} or active forces \cite{Manneville1999} to investigate couplings between  topography and drag. Also, one can imagine tuning flow boundary conditions of liquid droplets by controlled addition of lipids or other surfactants, spanning the range between liquid-like and solid-like behavior. Finally, we note that the methods presented here will also be applicable in non-Newtonian fluids, for which detailed understanding of microscale rheology continues to be an active area of study.

{\em Acknowledgements}. This material is based in part upon work supported by the National Science Foundation under Award Number 1507115. Any opinions, findings, and conclusions or recommendations expressed in this material are those of the author(s) and do not necessarily reflect the views of the National Science Foundation.

\bibliography{VesicleDiffusionBib}
\bibliographystyle{apsrev}

\end{document}